\documentclass[10pt,fleqn]{article}
\usepackage[utf8]{inputenc}
\usepackage[top=.75in,bottom=.75in,left=.75in,right=.75in]{geometry}
\usepackage{fouriernc}
\usepackage{authblk}
\usepackage{multicol}
\usepackage{graphicx}
\usepackage{amsmath}
\usepackage{amssymb}
\usepackage{titlesec}
\titleformat*{\section}{\large\bfseries}
\newcommand{\ket}[1]{\rvert#1\rangle}
\newcommand{\bra}[1]{\langle #1\rvert}
\newcommand{\braket}[2]{\langle #1\rvert#2\rangle}

\newcommand{\kket}[2]{\rvert#1\rangle_{#2}}

\newcommand{\Braket}[3]{\bra{#1}#2\ket{#3}}
\newcommand{\BBraket}[5]{{}_{#1}\bra{#2}\,#3\,\ket{#4}_{#5}}
\title{Spin lattices, state transfer and bivariate Krawtchouk polynomials}
\author[1]{Vincent X. Genest}
\author[2]{Hiroshi Miki}
\author[1]{Luc Vinet}
\author[3]{Alexei Zhedanov}
\affil[1]{Centre de recherches math\'ematiques, Universit\'e de Montr\'eal, P.O. Box 6128, Centre-ville Station, Montr\'eal, Canada, H3C 3J7}
\affil[2]{Department of Electronics, Doshisha University, Kyoto 6100394, Japan}
\affil[3]{Donetsk Institute for Physics and Technology, Donetsk 340114, Ukraine}
\date{}
\begin{document}
\maketitle
\hrule
\begin{abstract}
\noindent
The quantum state transfer properties of a class of two-dimensional spin lattices on a triangular domain are investigated. Systems for which the 1-excitation dynamics is exactly solvable are identified. The exact solutions are expressed in terms of the bivariate Krawtchouk polynomials that arise as matrix elements of the unitary representations of the rotation group on the states of the three-dimensional harmonic oscillator.
\end{abstract}
\hrule
\begin{multicols}{2}
\section{Introduction}
\noindent
The transfer of quantum states between distant locations is an important task in quantum information processing \cite{Bose-2007,Kay-2010}. To perform this task, one needs to design quantum devices that effect this transfer, i.e. devices such that an input state at one location is produced as output state at another location. A desirable property is that the transfer be realized with a high fidelity. When the input state is recovered with probability $1$, one has perfect state transfer (PST). One idea to attain perfect state transfer is to exploit the intrinsic dynamics of quantum systems so as to minimize the need of external controls and reduce noise.

Dynamical PST can for instance be achieved using one-dimen\-sion\-al spin chains \cite{Albanese-2004}. In the simplest examples, one considers chains consisting of $N+1$ spins with states
\begin{align*}
\ket{1}=
\begin{pmatrix}
1 \\ 0
\end{pmatrix},\quad 
\ket{0}=
\begin{pmatrix}
0 \\ 1
\end{pmatrix},
\end{align*}
and nearest-neighb\-or non-homogeneous couplings. These spin chains are governed by Hamiltonians of the form
\begin{equation}
\label{1D-Hamiltonian}
 H=\sum_{i=0}^{N}\left[\frac{J_{i+1}}{2}\big(\sigma_{i}^{x}\sigma_{i+1}^{x}+\sigma_{i}^{y}\sigma_{i+1}^{y}\big)+\frac{B_i}{2}\big(\sigma_i^{z}+1\big)\right],
\end{equation}
where $\sigma_{i}^{x}$, $\sigma_i^{y}$ and $\sigma_{i}^{z}$ are the Pauli matrices
\begin{gather*}
\sigma^{x}=
\begin{pmatrix}
0 & 1 \\
1 & 0
\end{pmatrix},\;
\sigma^{y}=
\begin{pmatrix}
0 & -i \\
i & 0
\end{pmatrix},\;
\sigma_{z}=
\begin{pmatrix}
1 & 0\\
0 & -1
\end{pmatrix},
\end{gather*}
acting on the spin located at the site $i$, where $i\in \{0,\ldots,N\}$. The coefficients $J_i$ are the coupling strengths between nearest neighbor sites and $B_i$ is the magnetic field strength at the site $i$. The state $\ket{0,\ldots,0}=\ket{0}^{\otimes (N+1)}$ is the ground state of $H$ with
\begin{align*}
H\ket{0}^{\otimes (N+1)}=0.
\end{align*}
The transfer properties of the chain defined by \eqref{1D-Hamiltonian} are exhibited as follows. Introduce the unknown state $\ket{\psi}=\alpha \ket{0}+\beta \ket{1}$ at the site $i=0$. One would like to recuperate $\ket{\psi}$ on the last site $i=N$ after some time. Since the component $\ket{0}^{\otimes (N+1)}$ is stationary, this amounts to finding the transition probability from the state $\ket{1}\otimes\ket{0}^{\otimes N}$ to the state $\ket{0}^{\otimes N}\otimes \ket{1}$. Thus, one only needs to consider the states with a single excitation; this can be done since the dynamics preserve the number of excitations. Perfect state transfer will be effected by the spin chains \eqref{1D-Hamiltonian} if there is a finite time $T$ such that 
\begin{align*}
U(T)\ket{1}\otimes \ket{0}^{\otimes N}=e^{i\phi}\ket{0}^{\otimes N}\otimes \ket{1},
\end{align*}
where $U(T)=e^{-i H}$. This is found to happen under appropriate choices of $J_{i}$ and $B_i$ \cite{VDJ-2010,Vinet_Zhedanov_2012,VZ-2012}.

Here we shall be concerned with the study of state transfer in two dimensions. We shall consider two-dimensional spin lattices with non-homogeneous nearest-neigh\-bor couplings on a triangular domain and identify the systems for which the 1-excitation dynamics is exactly solvable and exhibits interesting quantum state transfer properties. This study will take us to introduce and characterize orthogonal polynomials in two discrete variables by looking at matrix elements of reducible representations of $O(3)$ on the states of the three-dimensional harmonic oscillator. These polynomials will be identified with the bivariate Krawtchouk polynomials \cite{Genest-2013-06} .

The outline of the paper is as follows. In section 2, the two-dimensional spin lattices are introduced and their 1-excitation dynamics is discussed. In section 3, the connection between representations of the rotation group on oscillator states and bivariate Krawtchouk polynomials is made explicit. In section 4, the recurrence relations of the bivariate Krawtchouk polynomials are derived and are shown to provide exact solutions of the 1-excitation dynamics of a particular class of spin lattices.  In section 5, the generating function of the bivariate Krawtchouk polynomials is derived and is used to study the transfer properties of the spin lattices. A short conclusion follows.
\section{Triangular spin lattices and one-excitation dynamics}\noindent
We consider a uniform two-dimensional lattice on a triangular domain \cite{Miki-2012, Miki-2012-03}. 
\begin{center}
\scalebox{.25}{\includegraphics{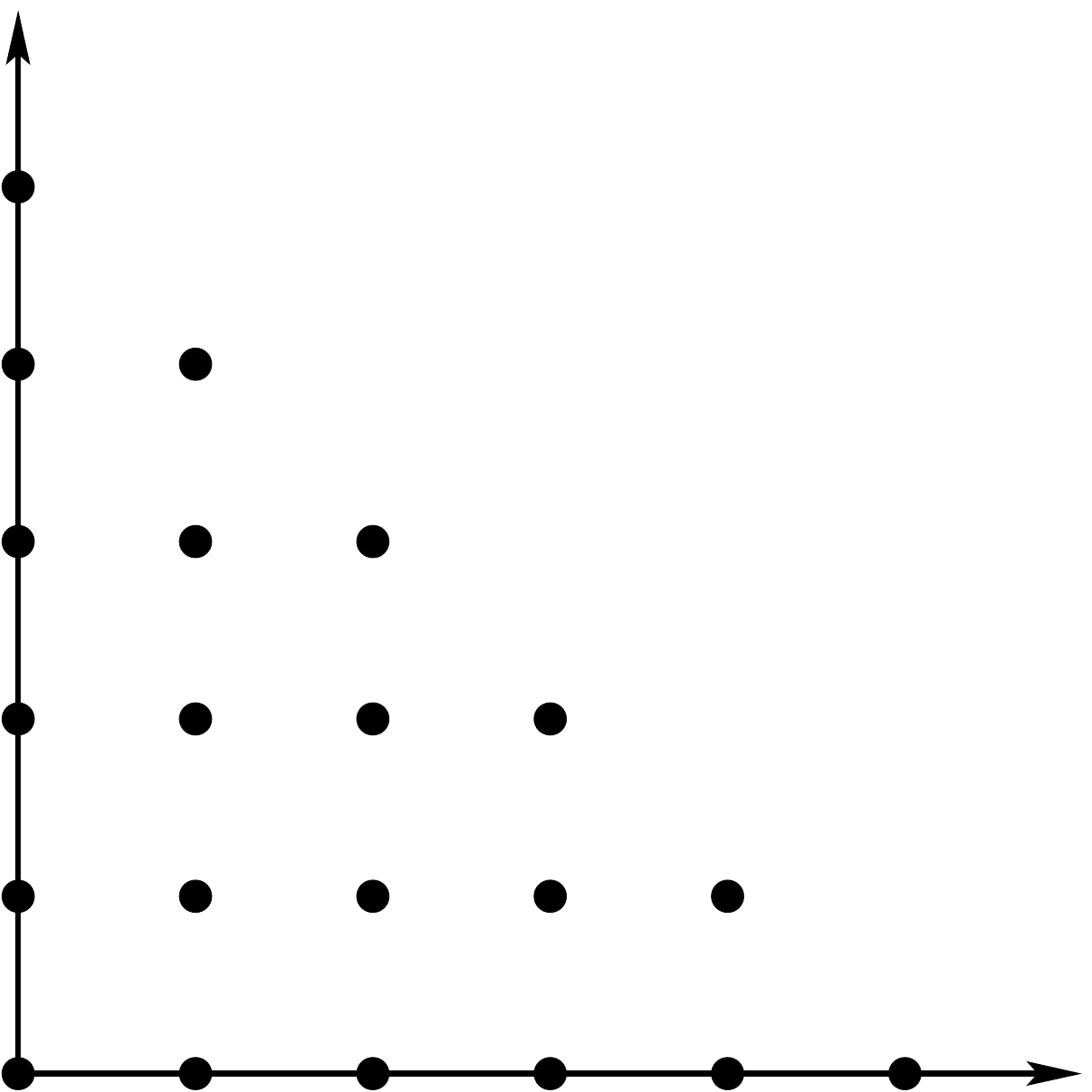}}
\\
Figure 1. Uniform two-dimensional lattice of triangular shape
\end{center}
The vertices of the lattice are labeled by the non-negative integers $(i,j)$ such that $i,j\in\{0,\ldots,N\}$ with $i+j\leq N$, where $N$ is also a non-negative integer. On each of the $(N+1)(N+2)/2$ sites of the lattice, there is a spin coupled to his nearest neighbors and to a local magnetic field. The Hamiltonian of the system is of the form 
\begin{multline}
\label{Spin-Lattice}
\mathcal{H}=\sum_{\substack{0\leq i,j\leq N\\ i+j\leq N}}\Bigg[\frac{I_{i+1,j}}{2}\big(\sigma_{i,j}^{x}\sigma_{i+1,j}^{x}+\sigma_{i,j}^{y}\sigma_{i+1,j}^{y}\big)\\
+\frac{J_{i,j+1}}{2}\big(\sigma_{i,j}^{x}\sigma_{i,j+1}^{x}+\sigma_{i,j}^{y}\sigma_{i,j+1}^{y}\big)+\frac{B_{i,j}}{2}\big(\sigma_{i,j}^{z}+1\big)\Bigg],
\end{multline}
where
\begin{align*}
I_{0,j}=J_{i,0}=0\;\text{and}\;I_{i,j}=J_{i,j}=0\;\text{if}\;i+j>N.
\end{align*}
The coefficients $I_{i,j}$ and $J_{i,j}$ are the coupling strengths between the sites $(i-1,j)$ and $(i,j)$ and between the sites $(i,j-1)$ and $(i,j)$, respectively. The total number of spins that are up (in state $\ket{1}$) over the lattice is a conserved quantity. Indeed, it is directly verified that 
\begin{align*}
\big[\mathcal{H}, \sum_{\substack{i,j\\ i+j\leq N}}\sigma_{i,j}^{z}\big]=0.
\end{align*}
Consequently, one can restrict the analysis of the Hamiltonian \eqref{Spin-Lattice} to the 1-excitation sector. A natural basis for the states of the lattice with only one spin up is provided by the vectors $\ket{i,j}$ labeled by the coordinates $(i,j)$ of the site where the spin up is located. One has thus
\begin{align*}
\ket{i,j}=E_{i,j},\quad i,j=0,\ldots, N,
\end{align*}
where $E_{i,j}$ is the $(N+1)\times (N+1)$ matrix that has a 1 in the $(i,j)$ entry and zeros everywhere else. The 1-excitation eigenstates of $\mathcal{H}$ are denoted by $\ket{x_{s,t}}$ and are defined by
\begin{align}
\label{E-Eigen}
\mathcal{H}\ket{x_{s,t}}=x_{s,t}\ket{x_{s,t}},
\end{align}
where $x_{s,t}$ is the energy eigenvalue. The expansion of the states $\ket{x_{s,t}}$  in the $\ket{i,j}$ basis is written as
\begin{align*}
\ket{x_{s,t}}=\sum_{\substack{0\leq i,j\leq N\\ i+j\leq N}} M_{i,j}(s,t) \ket{i,j}.
\end{align*}
Since both bases $\ket{x_{s,t}}$ and $\ket{i,j}$ are orthonormal, the transition matrix $M_{i,j}(s,t)$ is unitary. The energy eigenvalue equation \eqref{E-Eigen} imposes that the expansion coefficients $M_{i,j}(s,t)$ satisfy the 5-term recurrence relation
\begin{multline}
\label{5-Recurrence}
x_{s,t} M_{i,j}(s,t)=I_{i+1,j}M_{i+1,j}(s,t)+J_{i,j+1}M_{i,j+1}(s,t)
\\
+B_{i,j}M_{i,j}(s,t)+I_{i,j}M_{i-1,j}(s,t)+J_{i,j}M_{i,j-1}(s,t).
\end{multline}
In the following, we shall identity systems specified by the coupling strengths $I_{i,j}$,  $J_{i,j}$, and $B_{i,j}$ for which the spectrum $x_{s,t}$ and coefficients $M_{i,j}(s,t)$ can be exactly determined.
\section{Representations of $O(3)$ on oscillator states and orthogonal polynomials}\noindent
Consider the eigenstates 
\begin{equation*}
\ket{n_1,n_2,n_3}=\ket{n_1}\otimes\ket{n_2}\otimes \ket{n_3},\quad n_1,n_2,n_3=0,1,\ldots,
\end{equation*}
of the three-dimensional isotropic oscillator Hamiltonian 
\begin{align*}
H_{\text{osc}}=a_1^{\dagger}a_1+a_{2}^{\dagger}a_2+a_{3}^{\dagger}a_3,
\end{align*}
with  $H_{\text{osc}}\ket{n_1,n_2,n_3}=N\ket{n_1,n_2,n_3}$ where the eigenvalue is $N=n_1+n_2+n_3$. Recall that 
\begin{align*}
a_{i}\ket{n_i}=\sqrt{n_i}\ket{n_i},\quad a_{i}^{\dagger}\ket{n_i}=\sqrt{n_i+1}\ket{n_i+1}.
\end{align*}
Consider $R\in O(3)$, a rotation matrix. Define $U(R)$ the unitary representation of $O(3)$ by
\begin{align}
\label{Representation}
\begin{aligned}
U(R) a_{i} U^{\dagger}(R)=\sum_{k=1}^{3}R_{ki}a_{k},
\\
U(R) a_i^{\dagger}U^{\dagger}(R)=\sum_{k=1}^{3}R_{ki}a_{k}^{\dagger}.
\end{aligned}
\end{align}
It is directly seen from the above that $U(RS)=U(R)U(S)$ for $R$ and $S$ in $O(3)$, as should be for a representation. Furthermore, one has $U(R)U^{\dagger}(R)=U^{\dagger}(R)U(R)=1$. The oscillator Hamiltonian $H_{\text{osc}}$ is obviously invariant under rotations, i.e.
\begin{align*}
U(R) H_{\text{osc}} U^{\dagger}(R)=H_{\text{osc}},
\end{align*}
and thus any rotation stabilizes the energy eigenspaces of $H_{\text{osc}}$. The basis vectors for the eigensubspaces of $H_{\text{osc}}$ with a fixed value of the energy $N$ denoted by
\begin{align*}
\kket{i,j}{N}=\ket{i,j,N-i-j},
\end{align*}
transform reducibly among themselves under the action of the rotations. Consider the matrix elements of $U(R)$ in the basis $\{\kket{i,j}{N}\; \rvert\; i,j=0,\ldots,N; i+j\leq N\}$. These matrix elements can be cast in the form
\begin{align}
\label{Matrix-Elements}
\BBraket{N}{s,t}{U(R)}{i,j}{N}=W_{s,t;N}P_{i,j}(s,t;N),
\end{align}
where $P_{0,0}(s,t;N)\equiv 1$ and $W_{s,t;N}=\BBraket{N}{s,t}{U(R)}{0,0}{N}$. When no confusion can arise, we shall drop the explicit dependence of $U(R)$ on $R$ to ease the notation.
\subsection{Calculation of $W_{s,t;N}$}\noindent 
Let us first calculate the amplitude $W_{s,t;N}$. To that end, consider the matrix element $\BBraket{N-1}{s,t}{Ua_1}{0,0}{N}$. One has on the one hand
\begin{align*}
\BBraket{N-1}{s,t}{Ua_1}{0,0}{N}=0.
\end{align*}
On the other hand, one can write
\begin{multline*}
\BBraket{N-1}{s,t}{Ua_1}{0,0}{N}=\BBraket{N-1}{s,t}{Ua_1U^{\dagger}U}{0,0}{N}\\
=R_{11}\,\sqrt{s+1}\,\BBraket{N}{s+1,t}{U}{0,0}{N}\\
+R_{21}\,\sqrt{t+1}\,\BBraket{N}{s,t+1}{U}{0,0}{N}
\\
+R_{31}\,\sqrt{N-s-t}\,\BBraket{N}{s,t}{U}{0,0}{N}.
\end{multline*}
Combining the two equations above, one obtains
\begin{multline*}
R_{11}\,\sqrt{s+1}\,W_{s+1,t;N}+R_{21}\,\sqrt{t+1}\, W_{s,t+1;N}\\
+ R_{31}\,\sqrt{N-s-t}\,W_{s,t;N}=0.
\end{multline*}
Similarly, using $\BBraket{N-1}{s,t}{U(R)a_{2}}{0,0}{N}=0$, one finds
\begin{multline*}
R_{12}\,\sqrt{s+1}\,W_{s+1,t;N}+R_{22}\,\sqrt{t+1}\, W_{s,t+1;N}\\
+ R_{32}\,\sqrt{N-s-t}\,W_{s,t;N}=0.
\end{multline*}
Recalling that $\sum_{k=1}^{3}R_{ks}R_{kt}=\delta_{st}$, i.e. that $W_{s,t;N}$ is ``essentially orthogonal'' to the 1\textsuperscript{st} and 2\textsuperscript{nd} column of $R$, one obtains
\begin{align*}
W_{s,t;N}=C\frac{R_{13}^{s}R_{23}^{t}R_{33}^{N-s-t}}{\sqrt{s!t!(N-s-t)!}}.
\end{align*}
The constant $C$ can be found from the normalization condition
\begin{align*}
1&=\BBraket{N}{0,0}{U^{\dagger}U}{0,0}{N}\\
&=\sum_{s+t\leq N}\BBraket{N}{0,0}{U^{\dagger}}{s,t}{N}\BBraket{N}{s,t}{U}{0,0}{N}
\\
&=\sum_{s+t\leq N}\rvert W_{s,t;N} \rvert^2,
\end{align*}
and the trinomial theorem
\begin{align*}
(x+y+z)^{N}=\sum_{i+j\leq N}\frac{N!}{i!j!(N-i-j)!}x^{i}y^{j}z^{N-i-j},
\end{align*}
giving $C=\sqrt{N!}$ and thus 
\begin{align}
\label{W}
W_{s,t;N}=\binom{N}{s,t}^{1/2}R_{13}^{s}R_{23}^{t}R_{33}^{N-s-t},
\end{align}
where
\begin{align*}
\binom{N}{s,t}=\frac{N!}{s!t!(N-s-t)!}.
\end{align*}
\subsection{Raising relations}
\noindent
One can show that the functions $P_{i,j}(s,t;N)$ appearing in the matrix elements \eqref{Matrix-Elements} are polynomials of the discrete variables $s$ and $t$. One can write
\begin{align*}
\BBraket{N}{s,t}{Ua_{1}^{\dagger}}{i,j}{N-1}=\sqrt{i+1}\,W_{s,t;N}\,P_{i+1,j}(s;t,N),
\end{align*}
and also
\begin{align*}
&\BBraket{N}{s,t}{Ua_{1}^{\dagger}}{i,j}{N-1}=\BBraket{N}{s,t}{Ua_{1}^{\dagger}U^{\dagger}U}{i,j}{N-1}
\\
&=\sum_{\ell=1}^{3}R_{\ell,1}\;\BBraket{N}{s,t}{a_{\ell}^{\dagger} U}{i,j}{N-1}.
\end{align*}
Using \eqref{Matrix-Elements} and \eqref{W}, the two equations above yield
\begin{multline*}
\sqrt{N(i+1)}\,P_{i+1,j}(s,t;N)=\frac{R_{11}}{R_{13}}\,s\,P_{i,j}(s-1,t;N-1)\\
+\frac{R_{21}}{R_{23}}\,t\,P_{i,j}(s,t-1;N-1)
\\
+\frac{R_{31}}{R_{33}}\,(N-s-t)\,P_{i,j}(s,t;N-1).
\end{multline*}
A similar relation is obtained starting instead from the matrix element $\BBraket{N}{s,t}{Ua_{2}^{\dagger}}{i,j}{N-1}$:
\begin{multline*}
\sqrt{N(j+1)}\,P_{i,j+1}(s,t;N)=\frac{R_{12}}{R_{13}}\,s\,P_{i,j}(s-1,t;N-1)\\
+\frac{R_{22}}{R_{23}}\,t\,P_{i,j}(s,t-1;N-1)
\\
+\frac{R_{32}}{R_{33}}\,(N-s-t)\,P_{i,j}(s,t;N-1).
\end{multline*}
The two equations above show that the functions $P_{i,j}(s,t;N)$ are polynomials of total degree $i+j$ in the two variables $s,t$. Indeed, they allow to construct the $P_{i,j}(s,t;N)$ step by step from $P_{0,0}=1$ by iterations that only involve multiplications by the variables $s$ and $t$.
\subsection{Orthogonality relation}\noindent
The fact that the polynomials $P_{i,j}(s,t;N)$ are orthogonal follows from the unitarity of the representation $U(R)$ and from the fact that the states $\kket{i,j}{N}$ are orthonormal. The relation
\begin{align*}
&\BBraket{N}{i',j'}{U^{\dagger}U}{i,j}{N}
\\
&=\sum_{s+t\leq N}\BBraket{N}{i',j'}{U^{\dagger}}{s,t}{N}\BBraket{N}{s,t}{U}{i,j}{N}=\delta_{ii'}\delta_{jj'},
\end{align*}
translates into
\begin{align*}
\sum_{\substack{0\leq s, t \leq N \\s+t\leq N}}\omega_{s,t;N}\,P_{i,j}(s,t;N)\,P_{i',j'}(s,t;N)=\delta_{ii'}\delta_{jj'}.
\end{align*}
Thus the $P_{i,j}(s,t;N)$ are polynomials of two discrete variables that are orthogonal on the finite grid $s+t\leq N$ with respect to the trinomial distribution
\begin{align*}
\omega_{s,t;N}=W_{s,t;N}^2=\binom{N}{s,t}R_{13}^{2s}R_{23}^{2t}R_{33}^{2(N-s-t)}.
\end{align*}
They provide a two-variable generalization of the one-variable Krawtchouk polynomials which are orthogonal with respect to the binomial distribution \cite{Koekoek-2010, Grunbaum-2011-12, Grunbaum-2007-05, Griffiths-1971-04, Diaconis-02-2014, Iliev-2010}.
\section{Recurrence relations and exact solutions of 1-excitation dynamics}\noindent
We shall now derive the recurrence relations satisfied by the polynomials $P_{i,j}(s,t;N)$ and compare them with \eqref{5-Recurrence}. Consider the matrix element $\BBraket{N}{s,t}{a_{1}^{\dagger}a_1U}{i,j}{N}$. One has
\begin{equation*}
\BBraket{N}{s,t}{a_{1}^{\dagger}a_1 U}{i,j}{N}=s\,\BBraket{N}{s,t}{U}{i,j}{N}.
\end{equation*}
Using \eqref{Representation}, one has also
\begin{multline*}
\BBraket{N}{s,t}{a_{1}^{\dagger}a_1 U}{i,j}{N}
\\
=\sum_{m,n=1}^{3}R_{1m}R_{1n}\;\BBraket{N}{s,t}{U a_{m}^{\dagger}a_{n}}{i,j}{N}.
\end{multline*}
Equating the RHS of the two above equations and using the expression \eqref{Matrix-Elements} for the matrix elements, one finds
\small
\begin{align}
\nonumber
&sP_{i,j}(s,t;N)=\left[R_{11}^2\,i+R_{12}^2\,j+R_{13}^2(N-i-j)\right]P_{i,j}(s,t;N)
\\
\nonumber
&+R_{11}R_{13}\Big[\alpha_{i+1,j}\,P_{i+1,j}(s,t;N)+\alpha_{i,j}\,P_{i-1,j}(s,t;N)\Big]\\
\label{Recu-1}
&+R_{12}R_{13}\Big[ \beta_{i,j+1}\,P_{i,j+1}(s,t;N)+\beta_{i,j}P_{i,j-1}(s,t;N)\Big]
\\
\nonumber
&+R_{11}R_{12}\Big[\gamma_{i,j+1}\,P_{i-1,j+1}(s,t;N)+\gamma_{i+1,j}\,P_{i+1,j-1}(s,t;N)\Big],
\end{align}
\normalsize
where
\begin{align*}
&\alpha_{i,j}=\sqrt{i(N-i-j+1)},\;\beta_{i,j}=\sqrt{j(N-i-j+1)},
\\
&\gamma_{i,j}=\sqrt{i\,j}.
\end{align*}
Proceeding likewise with $\BBraket{N}{s,t}{a_{2}^{\dagger}a_2U}{i,j}{N}$, one obtains
\small
\begin{align}
\nonumber
&tP_{i,j}(s,t;N)=\left[R_{21}^2\,i+R_{22}^2\,j+R_{23}^2(N-i-j)\right]P_{i,j}(s,t;N)
\\
\nonumber
&+R_{21}R_{23}\Big[\alpha_{i+1,j}P_{i+1,j}(s,t;N)+\alpha_{i,j}P_{i-1,j}(s,t;N)\Big]\\
\label{Recu-2}
&+R_{22}R_{23}\Big[ \beta_{i,j+1}P_{i,j+1}(s,t;N)+\beta_{i,j}P_{i,j-1}(s,t;N)\Big]
\\
\nonumber
&+R_{21}R_{22}\Big[\gamma_{i,j+1}P_{i-1,j+1}(s,t;N)+\gamma_{i+1,j}P_{i+1,j-1}(s,t;N)\Big].
\end{align}
\normalsize
Upon combining the recurrence relations \eqref{Recu-1} and \eqref{Recu-2}, one can eliminate the non nearest-neighbor terms $P_{i-1,j+1}(s,t;N)$ and $P_{i+1,j-1}(s,t;N)$ to find
\small
\begin{multline*}
(R_{21}R_{22}s-R_{11}R_{12}t)P_{i,j}(s,t;N)=
\\
\Big\{ [R_{21}R_{22}(R_{11}^2-R_{13}^2)-R_{11}R_{12}(R_{21}^2-R_{23}^2)]\,i
\\
+[R_{21}R_{22}(R_{12}^2-R_{13}^2)-R_{11}R_{12}(R_{22}^2-R_{23}^2)]\,j
\\
+[R_{21}R_{22}R_{13}^2-R_{11}R_{12}R_{23}^2] N\Big\} P_{i,j}(s,t;N)\\
+\Big\{R_{21}R_{22}R_{11}R_{13}-R_{11}R_{12}R_{21}R_{23}\Big\}
\\
\times \Big[\alpha_{i,j}P_{i-1,j}(s,t;N)+\alpha_{i+1,j}P_{i+1,j}(s,t;N)\Big]\\
+\Big\{R_{21}R_{22}R_{12}R_{13}-R_{11}R_{12}R_{22}R_{23}\Big\}
\\
\times \Big[\beta_{i,j}\,P_{i,j-1}(s,t;N)+\beta_{i,j+1}\,P_{i,j+1}(s,t;N)\Big].
\end{multline*}
\normalsize
It is readily noted that the above relation is of the same form as the 5-term recurrence equation \eqref{5-Recurrence} that one has to solve to obtain the 1-excitation dynamics of the spin lattices governed by the Hamiltonian \eqref{Spin-Lattice}. Take
\begin{align}
\begin{aligned}
\label{Coef-1}
& I_{i,j}=(R_{21}R_{22}R_{11}R_{13}-R_{11}R_{12}R_{21}R_{23})\alpha_{i,j},
\\
& J_{i,j}=(R_{21}R_{22}R_{12}R_{13}-R_{11}R_{12}R_{22}R_{23})\beta_{i,j},
\end{aligned}
\end{align}
and
\begin{multline}
\label{Coef-2}
B_{i,j}=\Big\{[R_{21}R_{22}(R_{11}^2-R_{13}^2)-R_{11}R_{12}(R_{21}^2-R_{23}^2)]i
\\
+[R_{21}R_{22}(R_{12}^2-R_{13}^2)-R_{11}R_{12}(R_{22}^2-R_{23}^2)]j
\\
+[R_{21}R_{22}R_{13}^2-R_{11}R_{12}R_{23}^2]N\Big\}.
\end{multline}
Our polynomial analysis shows that the spectrum of the Hamiltonian \eqref{Spin-Lattice} with couplings \eqref{Coef-1}, \eqref{Coef-2} is given by
\begin{align*}
x_{s,t}=R_{21}R_{22}s-R_{11}R_{12}t,\quad s,t\in \{0,\ldots, N\},
\end{align*}
with $s+t\leq N$and that the unitary expansion coefficients are
\begin{align*}
M_{i,j}(s,t)=\BBraket{N}{s,t}{U(R)}{i,j}{N}=W_{s,t;N}\,P_{i,j}(s,t;N).
\end{align*}
The rotation matrix elements $R_{ij}$ are parameters. If one takes for instance
\begin{align}
\label{Rotation}
R=
\begin{pmatrix}
\frac{1}{2}-\frac{\sqrt{2}}{4} & -\frac{1}{2}-\frac{\sqrt{2}}{4} & \frac{1}{2} \\
-\frac{1}{2}-\frac{\sqrt{2}}{4} & \frac{1}{2}-\frac{\sqrt{2}}{4}  & \frac{1}{2}\\
\frac{1}{2} & \frac{1}{2} & \frac{\sqrt{2}}{2}
\end{pmatrix},
\end{align}
one has in particular 
\begin{align*}
R_{21}R_{22}=R_{11}R_{12}=-\frac{1}{8},\quad R_{13}=R_{23}=\frac{1}{2},
\end{align*}
and
\begin{align*}
&x_{s,t}=\frac{1}{8}(t-s),\quad &I_{i,j}=-\frac{1}{16}\sqrt{i(N-i-j+1)},
\\
&B_{i,j}=\frac{-1}{8\sqrt{2}}(j-i),\quad & J_{i,j}=\frac{1}{16}\sqrt{j(N-i-j+1)}.
\end{align*}
Note that the rotation $R$ specified by \eqref{Rotation} is improper since $\det R=-1$.
\section{State transfer}\noindent
Knowing the 1-excitation dynamics for the particular class of spin lattices, one can determine the transition amplitudes. Let $f_{(i,j), (k,\ell)}(T)$ denote the transition amplitude for the excitation at site $(i,j)$ to be found at the site $(k,\ell)$ after some time $T$. One can write
\begin{align*}
&f_{(i,j), (k,\ell)}(T)=\Braket{i,j}{e^{-iT\mathcal{H}}}{k,\ell}
\\
&=\sum_{s+t\leq N}\Braket{i,j}{e^{-iT\mathcal{H}}}{x_{s,t}}\braket{x_{s,t}}{k,\ell}
\\
&=\sum_{s+t\leq N}M_{i,j}(s,t)\,M_{k,\ell}(s,t)\,e^{-iT x_{s,t}}
\\
&=\sum_{s+t\leq N}\BBraket{N}{s,t}{U(R)}{i,j}{N}\,\BBraket{N}{s,t}{U(R)}{k,\ell}{N}\,e^{-iTx_{s,t}},
\end{align*}
with $x_{s,t}=R_{21}R_{22}s-R_{11}R_{12}t$. Typically one wishes to transfer state from a given site taken to be $(0,0)$. Using the expression \eqref{W} for $W_{s,t;N}$, the transition amplitude from the site $(0,0)$ to an arbitrary site $(i,j)$ is seen to be of the form
\begin{multline*}
f_{(0,0), (i,j)}=R_{33}^{N}\sum_{s+t\leq N}
\\
\sqrt{\binom{N}{s,t}}\left(\frac{R_{13}z_1}{R_{33}}\right)^{s}\left(\frac{R_{23}z_2}{R_{33}}\right)^{t}\BBraket{N}{s,t}{U(R)}{i,j}{N}
\end{multline*}
where we have taken
\begin{align}
\label{Z}
z_1=e^{-iR_{21}R_{22} T},\quad  z_2=e^{iR_{11}R_{12}T}.
\end{align}
Introduce another variable $u$ such that $s+t+u=N$ as well as an auxiliary variable $z_3$. Let
\begin{align*}
\alpha_1=R_{13}z_1,\quad \alpha_2=R_{23}z_2,\quad \alpha_3=R_{33}z_3.
\end{align*}
and define
\begin{multline}
\label{Gen-Fun}
G_{i,j;N}(\alpha_1,\alpha_2,\alpha_3)
\\
=\sum_{\substack{s,t,u\\ s+t+u=N}}\sqrt{\frac{N!}{s!t!u!}}\,\Braket{s,t,u}{U(R)}{i,j,k}\,\alpha_1^{s}\alpha_2^{t}\alpha_3^{u},
\end{multline}
with $i+j+k=N$. It is seen that $G_{i,j;N}(\alpha_1,\alpha_2,\alpha_3)$ is a generating function for $\BBraket{N}{s,t}{U(R)}{i,j}{N}$ and that
\begin{align}
\label{Amplitude-Gen}
f_{(0,0),(i,j)}=G_{i,j;N}(R_{13}z_1,R_{23}z_2,R_{33})\qquad z_3=1.
\end{align}
The generating function $G_{i,j;N}(\alpha_1,\alpha_2,\alpha_3)$ is readily com\-puted in the representation framework. Using \eqref{Gen-Fun}, one writes
\begin{align*}
&G_{i,j;N}(\alpha_1,\alpha_2,\alpha_3)=\sqrt{N!}
\\
&\times \sum_{s+t+u=N}\Braket{0,0,0}{\frac{(\alpha_1 a_1)^{s}}{s!} \frac{(\alpha_2 a_2)^{t}}{t!} \frac{(\alpha_3 a_3)^{u}}{u!}\,U}{i,j,k}
\\
&=\sqrt{N!}\;\Braket{0,0,0}{UU^{\dagger}e^{(\alpha_1 a_1+\alpha_2 a_2+\alpha_3 a_3)}U}{i,j,k},
\end{align*}
since $U$ keeps $N$ fixed and since the states are orthonormal. Because $U\ket{0,0,0}=\ket{0,0,0}$ and
\begin{align*}
U^{\dagger}e^{\sum_{\ell} \alpha_{\ell}a_{\ell}}U=e^{\sum_{\ell}\alpha_{\ell} Ua_{\ell}U^{\dagger}}=e^{\sum_{p}\beta_{p}a_{p}}
\end{align*}
with $\beta_{p}=\sum_{\ell}R_{\ell p}\alpha_{\ell}$, one can write
\begin{align*}
&G_{i,j;N}(\alpha_1,\alpha_2,\alpha_3)
\\
&=\sqrt{N!}\Braket{0,0,0}{e^{\beta_1 a_1+\beta_2 a_2+\beta_3 a_3}}{i,j,k}
\\
&=\sqrt{N!}\sum_{\ell,m,n}\frac{\beta_1^{\ell}\beta_2^{m}\beta_{3}^{n}}{\sqrt{\ell!m!n!}}\braket{\ell,m,n}{i,j,k},
\end{align*}
which gives
\begin{align*}
G_{i,j;N}(\alpha_1,\alpha_2,\alpha_3)=\binom{N}{i,j}^{1/2}\beta_1^{i}\beta_2^{j}\beta_{3}^{N-i-j},
\end{align*}
since $i+j+k=N$. Consequently, we have
\begin{multline*}
G_{i,j;N}(\alpha_1,\alpha_2,\alpha_3)=\sqrt{\binom{N}{i,j}}(R_{11}\alpha_1+R_{21}\alpha_2+R_{31}\alpha_3)^{i}
\\
\times (R_{12}\alpha_1+R_{22}\alpha_{2}+R_{32}\alpha_3)^{j}
\\
\times (R_{13}\alpha_1+R_{23}\alpha_{2}+R_{33}\alpha_3)^{N-i-j}.
\end{multline*}
In view of \eqref{Amplitude-Gen}, we have obtained the following formula for the transition amplitude
\begin{multline*}
f_{(0,0), (i,j)}(T)=
\\
\sqrt{\binom{N}{i,j}}(R_{11}R_{13}z_1+R_{21}R_{23}z_2+R_{31}R_{33})^{i}
\\
\times (R_{12}R_{13}z_1+R_{22}R_{23}z_2+R_{32}R_{33})^{j}
\\
\times (R_{13}^2z_1+R_{23}^2z_2+R_{33}^2)^{N-i-j},
\end{multline*}
with $z_1$ and $z_2$ given by \eqref{Z}. Let $R_{21}R_{22}=R_{11}R_{12}$ and take $T=\frac{\pi}{R_{11}R_{12}}$ so that $z_1=z_2=-1$. We have
\begin{multline*}
f_{(0,0), (i,j)}\left(\frac{\pi}{R_{11}R_{12}}\right)=\sqrt{\binom{N}{i,j}}
\\
\times (-R_{11}R_{13}-R_{21}R_{23}+R_{31}R_{33})^{i}
\\
\times (-R_{12}R_{13}-R_{22}R_{23}+R_{32}R_{33})^{j}
\\
\times (-R_{13}^2-R_{23}^2+R_{33}^2)^{N-i-j}.
\end{multline*}
If one adds to $R_{21}R_{22}=R_{11}R_{12}$ the condition $R_{33}=\sqrt{2}/2$, this implies that $f_{(0,0), (i,j)}\left(\frac{\pi}{R_{11}R_{12}}\right)=0$ unless $i+j=N$ since $(-R_{13}^2-R_{23}^2+R_{33}^2)=0$. These conditions were met by the rotation matrix considered in \eqref{Rotation}. With these conditions, the amplitude reads
\begin{multline*}
f_{(0,0), (i,j)}\left(\frac{\pi}{R_{11}R_{12}}\right)
\\
=\sqrt{\binom{N}{i,j}} (\sqrt{2}R_{31})^{i}(\sqrt{2}R_{32})^{j}\delta_{i+j,N},
\end{multline*}
and the output excitation distributes binomially on the site of the boundary hypotenuse. Hence for the values of the parameters such that $R_{21}R_{22}=R_{11}R_{12}$ and $R_{33}=\sqrt{2}/2$, the Hamiltonian $\mathcal{H}$ with non-homogeneous couplings \eqref{Coef-1} and \eqref{Coef-2} will dynamically evolve the state $\ket{0,0}$ in time $\frac{\pi}{R_{11}R_{12}}$ to any one of the states $\ket{i,N-i}$ with probability $1$. As a consequence
\begin{align*}
\Big\rvert f_{(0,0), (i,j)}\left(\frac{\pi}{R_{11}R_{12}}\right)\Big\rvert^2=0,\;\text{when}\; i+j<N,
\end{align*}
which is akin to perfect transfer. It can be shown that the bivariate Krawtchouk polynomials are symmetric for these values of the parameters \cite{Miki-2012}.

\section{Conclusion}\noindent
We have shown that the solutions of the 1-excitation dynamics for a particular class of spin networks with inhomogeneous couplings is tied to multivariate orthogonal polynomials and we have provided an illustration of the theory of multivariate Krawtchouk polynomials based on the  representations of $O(n)$ on oscillator states. For more details on the connection between orthogonal polynomials and perfect state transfer, the reader may wish to consult \cite{VZ-2012, Miki-2012,Post-2014}. For a detailed account of  the relation between multivariate orthogonal polynomials and Lie group representations, the reader is referred to \cite{Genest-2013-06,Genest-2013-07-2, Genest-2014-01,Genest-2014-05}.


\begin{thebibliography}{10}

\bibitem{Albanese-2004}
C.~Albanese, M.~Christandl, N.~Datta, and A.~Ekert.
\newblock {Mirror Inversion of Quantum States in Linear Registers}.
\newblock {\em Phys. Rev. Lett.}, 93:230502, 2004.

\bibitem{Bose-2007}
S.~Bose.
\newblock {Quantum communication through spin chain dynamics: an introductory
  overview}.
\newblock {\em Contemp. Phys.}, 48:13--30, 2007.

\bibitem{VDJ-2010}
R.~Chakrabarti and J.~Van der Jeugt.
\newblock {Quantum communication through a spin chain with interaction
  determined by a Jacobi matrix}.
\newblock {\em J. Phys. A: Math. Theor.}, 438:085302, 2010.

\bibitem{Diaconis-02-2014}
P.~Diaconis and R.~C. Griffiths.
\newblock {An introduction to multivariate Krawtchouk polynomials and their
  applications}.
\newblock {\em J. Stat. Plann. Inf.}, 2014.

\bibitem{Genest-2014-05}
V.~X. Genest, H.~Miki, L.~Vinet, and A.~Zhedanov.
\newblock {The multivariate Charlier polynomials as matrix elements of the
  Euclidean group representation on oscillator states}.
\newblock {\em J. Phys. A: Math. Theor.}, 47:215204, 2014.

\bibitem{Genest-2014-01}
V.~X. Genest, H.~Miki, L.~Vinet, and A.~Zhedanov.
\newblock {The multivariate Meixner polynomials as matrix elements of $SO(d,1)$
  representations on oscillator states}.
\newblock {\em J. Phys. A: Math. Theor.}, 47:045207, 2014.

\bibitem{Genest-2013-06}
V.~X. Genest, L.~Vinet, and A.~Zhedanov.
\newblock {The multivariate Krawtchouk polynomials as matrix elements of the
  rotation group representations on oscillator states}.
\newblock {\em J. Phys. A: Math. Theor.}, 46:505203, 2013.

\bibitem{Genest-2013-07-2}
V.~X. Genest, L.~Vinet, and A.~Zhedanov.
\newblock { Interbasis expansions for the isotropic 3D harmonic oscillator and
  bivariate Krawtchouk polynomials}.
\newblock {\em J. Phys. A: Math. Theor.}, 47:025202, 2014.

\bibitem{Griffiths-1971-04}
R.~C. Griffiths.
\newblock {Orthogonal Polynomials on the Multinomial Distribution}.
\newblock {\em Aus. J. Stat.}, 13:27--35, 1971.

\bibitem{Grunbaum-2007-05}
A.~Gr{\"u}nbaum.
\newblock {The Rahman Polynomials Are Bispectral}.
\newblock {\em SIGMA}, 302:65--75, 2007.

\bibitem{Grunbaum-2011-12}
A.~Gr{\"u}nbaum and M.~Rahman.
\newblock {A System of Multivariable Krawtchouk polynomials and a Probabilistic
  Application}.
\newblock {\em SIGMA}, 7:119--135, 2011.

\bibitem{Iliev-2010}
P.~Iliev and P.~Terwilliger.
\newblock {The Rahman polynomials and the Lie algebra $sl_{3}(\mathbb{C})$}.
\newblock {\em Trans. Amer. Math. Soc.}, 364:4225--4238, 2012.

\bibitem{Kay-2010}
A.~Kay.
\newblock {Perfect, efficient, state transfer and its application as a
  constructive tool}.
\newblock {\em Int. J. Qtm. Inf.}, 8:641--676, 2010.

\bibitem{Koekoek-2010}
R.~Koekoek, P.A. Lesky, and R.F. Swarttouw.
\newblock {\em {Hypergeometric orthogonal polynomials and their
  $q$-analogues}}.
\newblock Springer, 1\textsuperscript{st} edition, 2010.

\bibitem{Miki-2012-03}
H.~Miki, S.~Post, L.~Vinet, and A.~Zhedanov.
\newblock {A superintegrable finite oscillator in two dimensions with SU(2)
  symmetry}.
\newblock {\em J. Phys. A: Math. Theor.}, 46:125207, 2012.

\bibitem{Miki-2012}
H.~Miki, S.~Tsujimoto, L.~Vinet, and A.~Zhedanov.
\newblock {Quantum-state transfer in a two-dimensional regular spin lattice of
  triangular shape}.
\newblock {\em Phys. Rev. A}, 85:062306, 2012.

\bibitem{Post-2014}
S.~Post.
\newblock {Quantum perfect state transfer in a 2D lattice}.
\newblock {\em Acta Appl. Math.}, 2014.

\bibitem{Vinet_Zhedanov_2012}
L.~Vinet and A.~Zhedanov.
\newblock {Almost perfect state transfer in quantum spin chains}.
\newblock {\em Phys. Rev. A}, 86:052319, 2012.

\bibitem{VZ-2012}
L.~Vinet and A.~Zhedanov.
\newblock {How to construct spin chains with perfect state transfer}.
\newblock {\em Phys. Rev. A}, 85:012323, 2012.

\end{thebibliography}

\end{multicols}

\end{document}